\documentclass{article}

\usepackage{PRIMEarxiv}

\usepackage[utf8]{inputenc} 
\usepackage[T1]{fontenc}    
\usepackage{hyperref}       
\usepackage{url}            
\usepackage{booktabs}       
\usepackage{amsfonts}       
\usepackage{nicefrac}       
\usepackage{microtype}      
\usepackage{lipsum}
\usepackage{fancyhdr}       
\usepackage{graphicx}       
\graphicspath{{media/}}     
\usepackage[affil-it]{authblk}
\usepackage{amsmath}
\usepackage{subfigure}
\usepackage{enumitem}

\def\SYS{SGSM}
\def\CMP{\scshape Compressor}
\def\MX{{\scshape Mixer}}

\title{SGSM: A Foundation-model-like Semi-generalist Sensing Model
}

\author[1,2,3]{Tianjian Yang}
\author[1,2,3]{Hao Zhou}
\author[1]{Shuo Liu}
\author[1,2,3]{Kaiwen Guo}
\author[1]{Yiwen Hou}
\author[4]{Haohua Du}
\author[5]{Zhi Liu}
\author[1,2,3]{Xiang-Yang Li}
\affil[1]{School of Computer Science and Technology, University of Science and Technology of China}
\affil[2]{CAS Key Laboratory of Wireless-Optical Communications, University of Science and Technology of China}
\affil[3]{Deqing Alpha Innovation Institute, Huzhou, Zhejiang, China}
\affil[4]{School of Cyber Science and Technology, Beihang University}
\affil[5]{Department of Computer and Network Engineering, The University of Electro-Communications, Tokyo 182-8585, Japan}

\begin{document}
\maketitle

\begin{abstract}
The significance of intelligent sensing systems is growing in the realm of smart services. These systems extract relevant signal features and generate informative representations for particular tasks. However, building the feature extraction component for such systems requires extensive domain-specific expertise or data. The exceptionally rapid development of foundation models is likely to usher in newfound abilities in such intelligent sensing. We propose a new scheme for sensing model, which we refer to as semi-generalist sensing model ({\SYS}). {\SYS} is able to semiautomatically solve various tasks using relatively less task-specific labeled data compared to traditional systems. Built through the analysis of the common theoretical model, {\SYS} can depict different modalities, such as the acoustic and Wi-Fi signal. Experimental results on such two heterogeneous sensors illustrate that {\SYS} functions across a wide range of scenarios, thereby establishing its broad applicability. In some cases, {\SYS} even achieves better performance than sensor-specific specialized solutions. Wi-Fi evaluations indicate a 20\% accuracy improvement when applying {\SYS} to an existing sensing model.
\end{abstract}

\keywords{Mobile computing \and Machine learning}

\section{Introduction}
Intelligent sensing systems have shown remarkable performance on many environmental perception (e.g., liquid recognition \cite{re-1-liq-1}, soil moisture estimation \cite{re-1-soil-1}, temperature monitoring \cite{re-1-mmwave-2}) and human activity (e.g., fall detection \cite{re-1-fall-1}, vital sign estimation \cite{intro-vital-1}, location tracking \cite{re-1-track-1}) tasks, becoming the core component of smart physical-related services, such as smart city and smart manufacturing. However, the current cost of designing intelligent sensing systems is relatively high since the models were designed to solve specific tasks with expensive expert knowledge~\cite{HASsurvey} or a substantial amount of domain-specific data~\cite{limubert}, one at a time.

Foundation models \cite{bommasani2021opportunities} -- the latest generation of artificial intelligence (AI) models -- are intuitively used to generalize the model for numerous downstream tasks, which are trained on large multimodal datasets. They can solve entirely new tasks which the models are never explicitly trained for.
Although the foundation models paradigm perform well in computer vision or natural language processing area, applying them in the intelligent sensing area is still challenging for two reasons.

First, \textbf{it is difficult to generate or access massive and diverse sensing datasets}. 
Massive high-quality data is crucial for foundation model applications, such as computer vision \cite{betker2023improving} and natural language processing \cite{bommasani2021opportunities}. However, this requirement is often unmet in the sensing field. To overcome this challenge, signal processing methods based on spectrum analysis are traditionally used, such as Discrete Fourier Transform (DFT), Discrete Wavelet Transform (DWT), and Hilbert-Huang Transform (HHT). Unfortunately, these methods focus more on general features and require considerable expertise to apply them to specific tasks. This lack of standardized guidelines makes it difficult to efficiently utilize these handcrafted methods and the datasets transformed by them in AI models.

Secondly, \textbf{the sensors work on complex principles and the sensing tasks are ever-changing. }
The task-specific machine learning models are largely developed for handling this, but most of the existing solutions lose the generality and have to re-train for different scenarios.
For example, a wearable exercise monitoring model may be trained on an IMU dataset where samples are annotated as human activity. This model can only predict the exercise situation, but the same dataset with different labels could carry out other diagnostic model for some common health problem, like pneumonia.

Fundamentally, the challenges come from the phenomenon of \textit{feature drifts}, wherein the extracted features associated with the data analysis method are not well-suited for subsequent tasks. 
Therefore, to alleviate such a phenomenon, the intuitive idea is to offer a scheme that guides the design of specific sensing systems by \textit{using the advantages of both handcrafted and machine learning approaches} – utilizing the data handling knowledge from ready-made signal processing methods and the task depicting ability from machine learning models.

Nevertheless, the realization of such a scheme is not obvious. Utilizing existing human knowledge to enhance foundation model performance remains an ongoing and unresolved subject. Furthermore, the intrinsic complexity associated with human's and models' contexts have not been thoroughly investigated \cite{humanmodel}. There is currently a lack of standardized guidelines for extracting data handling knowledge from signal processing methods, and experience in generalizing the ability of task-specific models in the field of signal processing. The scheme must not only effectively incorporate both components, but also facilitate interactive and iterative adjustments between them that are both uncertain and subject to change.

In this paper, we propose {\SYS}, a novel intelligent sensing scheme to realize versatile models with limited labelled data and little signal processing knowledge.
It works in a foundation-model-style, i.e., a \textit{generalized} model to support numerous \textit{specialized} downstream tasks.
To achieve this target, {\SYS} utilizes a two-phase autoencoder (AE) structure to combine the knowledge and annotated data.
In the first phase, {\SYS} provides \textit{generalized} representations by extracting the knowledge of existing signal processing methods with a series of AEs denoted as {\CMP}s. They unify the outputs of diverse handcrafted methods into a shared feature space and learn to generate task-irrelative representations with unlabelled data.
In the second phase, {\SYS} supports \textit{specialized} tasks by guiding the selection of methods with a denoising AE denoted as the {\MX} which can automatically evaluate the performance of various method combinations regarding to a specific downstream task, using the representations of the annotated data from the {\CMP}s.

The main contributions of the paper are:
\begin{itemize}
    \item To the best of our knowledge, {\SYS} is the first foundation-like model to leverage the handcrafted knowledge and machine learning approaches in the intelligent sensing area, requiring only restricted annotated data: in this way we can achieve the same accuracy as existing works with zero expert knowledge and less amount of annotated training data.
    \item We propose a semi-automated intelligent sensing paradigm which can adaptively construct the sensing model, whatever the sensor is, whatever the sensing objectives are. {\SYS} does not base on particular characteristics derived from signal types, which supports the robustness of the scheme.
    \item We build prototypes of {\SYS} and test them with real world datasets. Experimental results on two heterogeneous sensors (acoustic and Wi-Fi) show that our scheme functions across a wide range of scenarios, thereby establishing its broad applicability. In some cases, {\SYS} even achieves better performance than sensor-specific specialized solutions. Specifically,  Wi-Fi evaluations indicate a 20\% accuracy improvement when applying {\SYS} to an existing sensing model.
\end{itemize}

\section{Methodology}\label{sec:method}
In this section, we will first give the problem formulation. Then, we will illustrate the basic idea of how {\SYS} works and provide illustrative case studies as support.

\subsection{Problem Formulation}\label{sec:problemform}
In the scenario of \textit{semi-generalist intelligent sensing}, a generic task consists of a sensing objective $o$, a sensor and its corresponding signal dataset $S$, and a related labeled dataset $D^o_S$.
It is noteworthy that $D^o_S$ has the same data type with $S$ and labeled for $o$.

Given an instance $(o, S,D^o_S)$ and the handcrafted signal processing methods $F = \{f_1, f_2, \cdots, f_n\}$ as input, our goal is to identify which subset $F' \subseteq F$ has the optimal performance with deep learning classification methods. To do so, we need to solve two co-related sub-problems. The first one is to find the optimal subset $F'$ on the dataset $D^o_S$,
\begin{equation}
    \phi(F', D^o_S)\geq \max_{\hat{F} \in \mathcal{P}(F)-F'} \left\{\phi\left(\hat{F},D^o_S\right)\right\} + \varepsilon,
    \label{eq:problem1}
\end{equation}
where $\phi$ itself can be interpreted as proper metrics such as accuracy and F1-score, and $\varepsilon$ is an error coefficient. 

The second is to guarantee the performance on the real task dataset $S$ being at least same with on $D^o_S$ for chosen $F'$,
 \begin{equation}
    |\phi(F', S) - \phi(F', D^o_S)| \leq \varsigma,
    \label{eq:problem2}
\end{equation}
where $\varsigma$ is an error coefficient. 
\subsection{Solution Sketch}
The key to solve problems (\ref{eq:problem1}) and (\ref{eq:problem2}) is to provide a proper way to evaluate the performance of $\hat{F}$ on $D_D^o$, which leads to two requirements. First, {\SYS} should be capable of normalizing the output of various signal processing methods so that it can deal with heterogeneous sequences which vary in scale and granularity. Second, {\SYS} should provide evaluation results of all method selections conveniently and cost-efficiently. 

\begin{figure}[htbp]
  \centering
  \includegraphics[width=0.65\columnwidth]{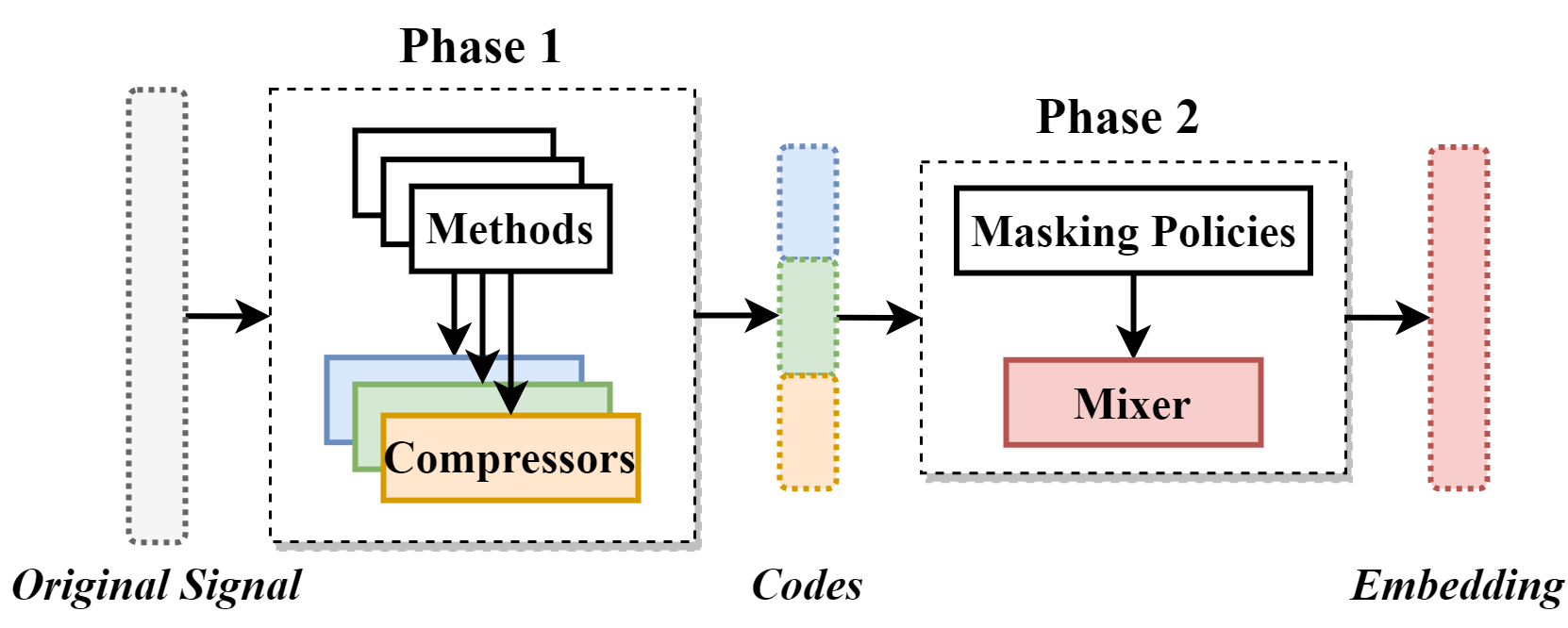}
  \caption{Two-phase structure of {\SYS}.}
  \label{fig:ideaFlow}
\end{figure}

The proposed system satisfies these requirements by applying a two-phase structure. The first phase, {\CMP}s, extracts compressed representations from transformed sequences generated by processing methods. {\CMP}s learn to standardize and refine given information, and are trained with unlabeled data to enhance robustness. The second phase, the {\MX}, combines representations from all method channels and integrates them into embeddings $\SYS(\hat{F},D^o_S)$, meanwhile providing a masking interface for subsequent evaluation.

\subsection{Phase One - {\CMP}s}
The first phase, {\CMP}s, focuses on dealing with task-irrelevant knowledge provided by existing signal processing methods. {\CMP}s extract information and normalize them for later use. 
{\CMP}s are a series of autoencoders, which have been part of the historical landscape of neural networks for decades and were traditionally used for dimensionality reduction or feature learning \cite{deeplearning}. In {\SYS}, signal processing methods are applied to unlabeled data $U$ for transformed results $f_i(U)$. {\CMP}s receive the results and generate the latent codes $\text{\CMP}(f_i(U))$. By learning from the transformed results of unlabeled data, {\CMP}s manage to refine the core knowledge of corresponding methods. Moreover, by processing heterogeneous information into standardized code vectors, {\CMP}s provide normalized preparation for later processing.

\subsection{Phase Two - {\MX}}

The second phase, the {\MX}, is responsible for using the refined codes provided by the {\CMP}s to generate final embeddings according to method selections. The embeddings of \textit{task-relevant} datasets can be used for subsequent evaluations.
To fulfill the {\MX}, we combine an autoencoder with masking policies, i.e., a masked autoencoder. Masking policies come from the need to adapt various method configurations $\hat{F}$. We regard researchers selecting methods as applying ``noise'' to corresponding channels of {\CMP}s. In this way, not using methods is equivalent to masking the related codes. 
With masking policies, the only challenge left is to combine information from masked codes. We apply masked autoencoders, denoising autoencoders (DAE) which receive corrupted data as input and predict the original, uncorrupted data as output. A DAE is intended not only to denoise but to learn a good internal representation as a side effect \cite{deeplearning}. If we regard the method configurations as noise, training a DAE with masking policies enables the DAE to deal with the ``noise'', i.e., varying method combinations, in advance. Thus, the system needs no extra training to generate embeddings for particular downstream data. In conclusion, by applying masked autoencoders, the system can combine information from different {\CMP} channels and generate embeddings with various method selections.

\section{System Design}\label{sec:system}
\begin{figure*}[htbp]
    \centering
    \includegraphics[width=\textwidth]{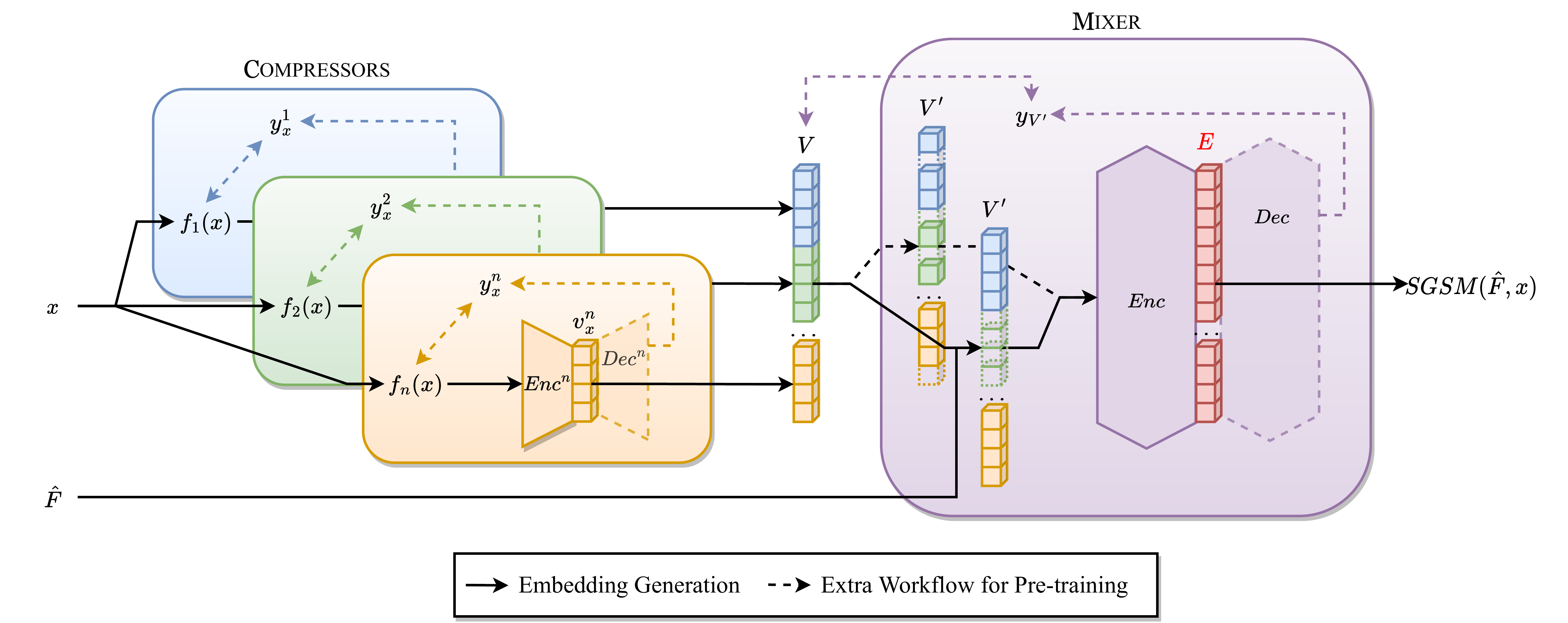}
    \caption{{\SYS}'s workflow for embedding generation and extra workflow for pre-training in {\SYS}.}
    \label{Workflow}
\end{figure*}

In this section, we will first describe design details of {\CMP}s and the {\MX}. After that, instructions on how to apply {\SYS} in specific tasks will be given. The overall workflow for embedding generation and extra workflow for pre-training are depicted in Fig. \ref{Workflow}.

\subsection{{\CMP}}
An instance of {\SYS} involves several {\CMP}s, each of which corresponds to a signal processing method. Without loss of generality, we assume that a set of methods $\left\{f_i|1\leq i\leq n\right\}$ is involved, where each method is a mapping $f_i:\mathbf{R}^L\rightarrow\mathbf{R}^{L_i'}$. $f_i(\boldsymbol{x})$ denotes the transformed output of a signal sequence $\boldsymbol{x}$ by $f_i$. The autoencoder of $f_i$ is composed of encoder $Enc^i$ and decoder $Dec^i$. For $f_i(\boldsymbol{x})$, $Enc^i$ generates a code of a fixed length:
\begin{equation}
\boldsymbol{v}_{\boldsymbol{x}}^i=Enc^i\left(f_i(\boldsymbol{x})\right),
\end{equation}
where $\boldsymbol{v}_{\boldsymbol{x}}^i\in\mathbf{R}^d$. The length is called fixed because all encoders share the same length of the codes. By this constraint, {\CMP}s transform sequences of all signal transformation results into a unified data space. Sequentially, the decoder $Dec^i$ accepts $\boldsymbol{v}^i_{\boldsymbol{x}}$ as input and outputs a vector of length $L_i'$:
\begin{equation}
\boldsymbol{y}_{\boldsymbol{x}}^i=Dec^i\left(\boldsymbol{v}_{\boldsymbol{x}}^i\right).
\end{equation}
The loss function of this autoencoder is given by:
\begin{equation}
    \mathcal{L}^i(\boldsymbol{x})=\frac{1}{L'_i}\left(\boldsymbol{y}_{\boldsymbol{x}}^i-f_i(\boldsymbol{x})\right)^2+1-\cos{\left(\boldsymbol{y}_{\boldsymbol{x}}^i, f_i(\boldsymbol{x})\right)}.
\end{equation}
The loss function can be considered as a combination of the mean squared error and the cosine embedding loss between the input and the reconstruction outcome. 
\begin{figure}[ht]
  \centering
  \includegraphics[width=0.6\columnwidth]{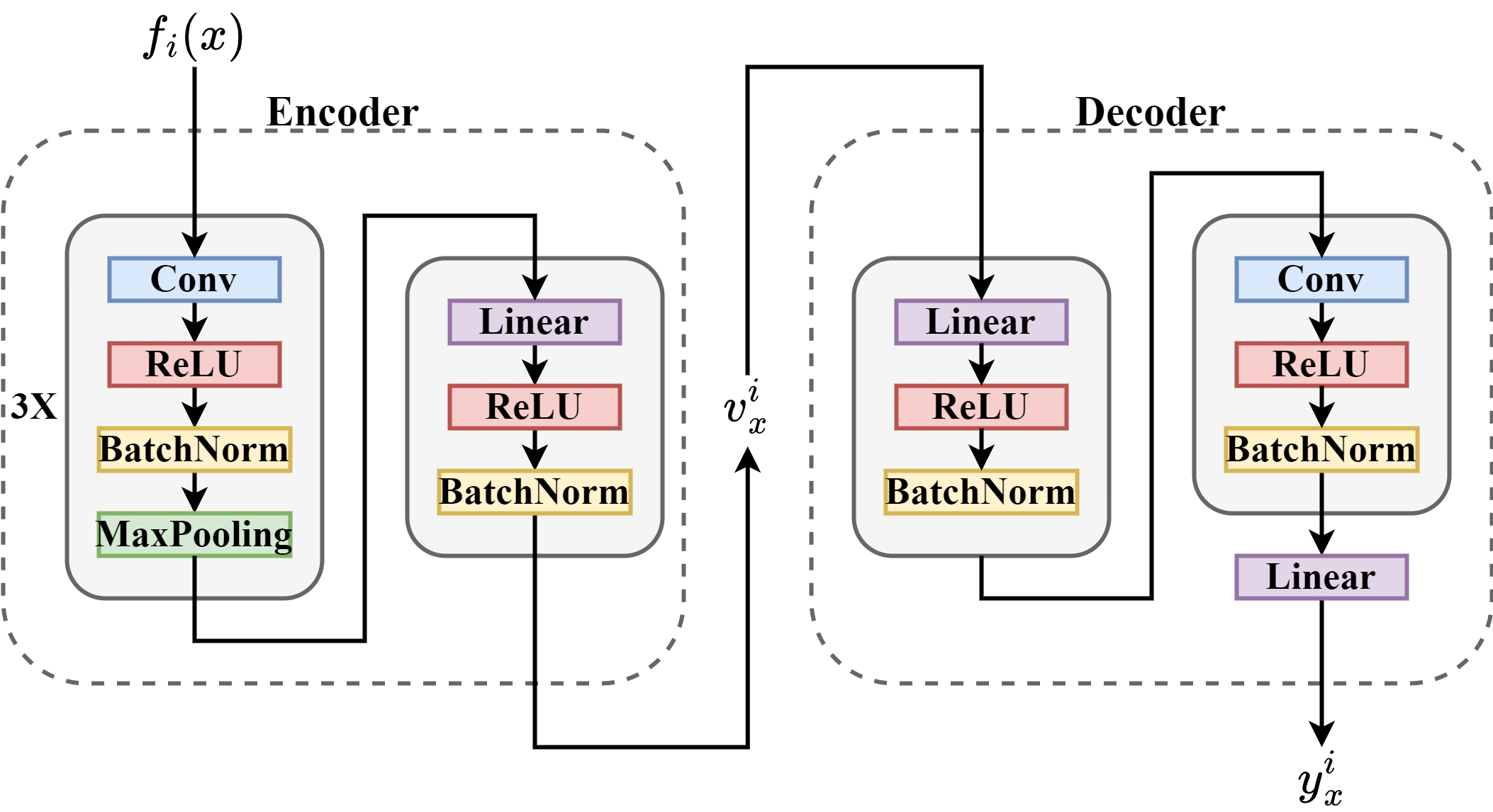}
  \caption{Network structure of {\CMP}.}
  \label{StrucCMP}
\end{figure}
\begin{figure}[ht]
  \centering
  \includegraphics[width=0.6\columnwidth]{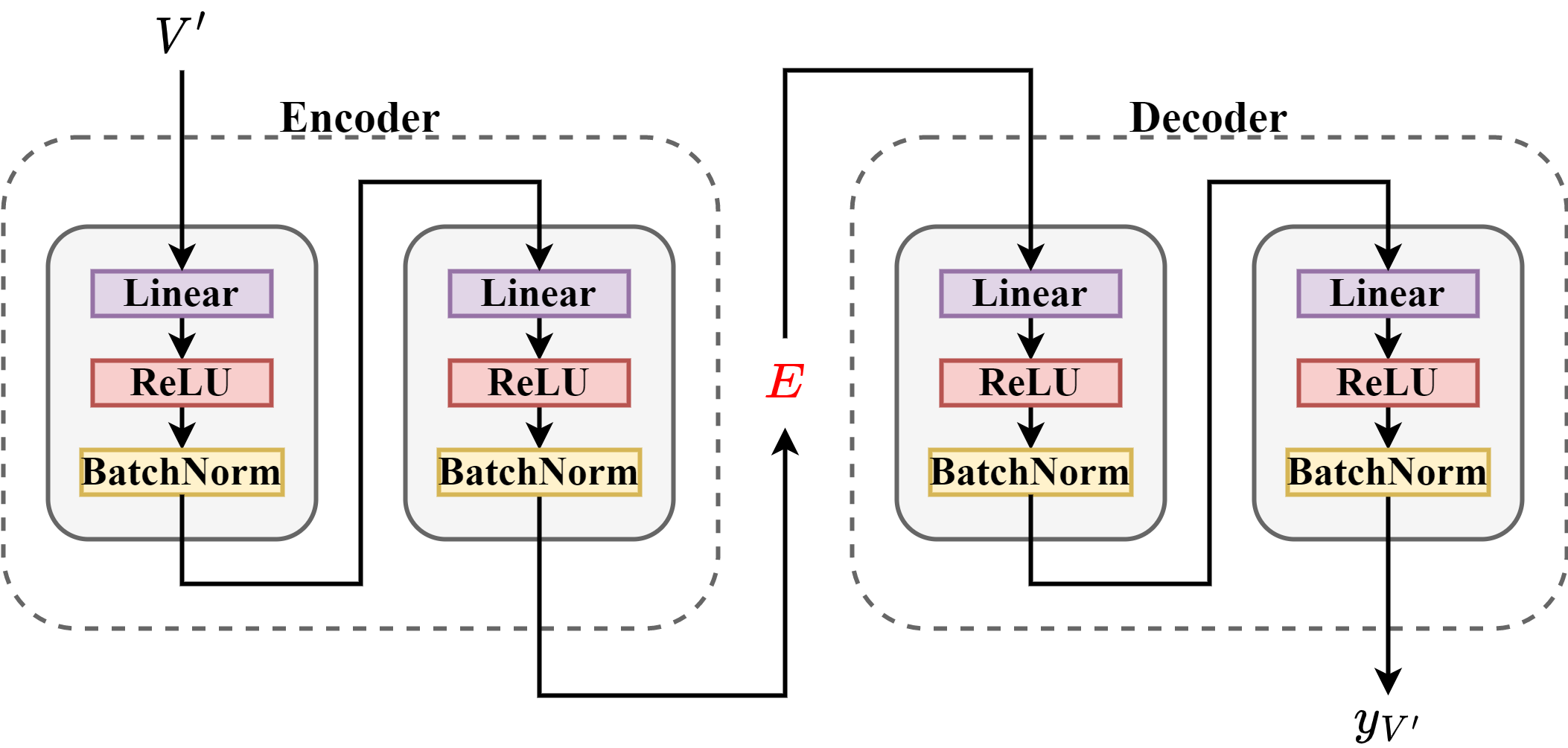}
  \caption{Network structure of {\MX}.}
  \label{StrucMX}
\end{figure}

Fig. \ref{StrucCMP} depicts the structure of the {\CMP}. Multiple layers are employed, including convolutional layers, ReLU layers, linear layers, etc. The encoder of {\CMP} consists of two parts. The first part aims to extract and refine information by increasing the number of channels. The second part aims to map the output of the first part to codes of a uniform length $d$. By assuring $d<L'_i$, the encoder must learn a compressed representation of input, which also explains why they are called {\CMP}s. 

The {\CMP} learns to compress and reconstruct transformed signal sequences based on the aforementioned loss function. The limitation of $d<L'_i$ requires the undercomplete AE to extract valuable information instead of plunging into the trivial solution of identity mapping.

\subsection{\MX}
There is one {\MX} in every instance of {\SYS}. Assume that a series of {\CMP}s are trained. For each raw signal sequence $\boldsymbol{x}$, {\CMP}s provide $n$ codes $\{\boldsymbol{v}_{\boldsymbol{x}}^i|1\leq i\leq n\}$. Concatenating these codes in a certain order results in a vector $\boldsymbol{V}$ of length $nd$. Let $\boldsymbol{V}=\boldsymbol{v}_{\boldsymbol{x}}^1\oplus \boldsymbol{v}_{\boldsymbol{x}}^2\oplus \cdots\oplus \boldsymbol{v}_{\boldsymbol{x}}^n$, where $\oplus$ indicates the concatenating operation. Applying one of the two masking policies produces a masked vector $\boldsymbol{V}'$.
\begin{itemize}
    \item Global-mask policy: $\boldsymbol{V}'$ is the outcome of setting a certain percentage of the bits of $\boldsymbol{V}$ to zero. In our experiments, this percentage is set to 10\%.
    \item Channel-mask policy: $\boldsymbol{V}'$ is the outcome of setting bits of randomly selected channels of $\boldsymbol{V}$ to zero. Each channel has a 50\% possibility to be masked, except that masking all channels is meaningless and thus forbidden. For example, if the third channel is selected when masking $\boldsymbol{V}=\boldsymbol{v}_{\boldsymbol{x}}^1\oplus \boldsymbol{v}_{\boldsymbol{x}}^2\oplus \boldsymbol{v}_{\boldsymbol{x}}^3$, then $\boldsymbol{V}'=\boldsymbol{v}_{\boldsymbol{x}}^1\oplus \boldsymbol{v}_{\boldsymbol{x}}^2\oplus\boldsymbol{0}$.
\end{itemize}
The possibility of the global-mask policy being applied is 80\% while that of the channel-mask policy is 20\%. While training the {\MX}, each vector $\boldsymbol{V}$ is masked independently every time it is fed into the {\MX}. These masked vectors $\boldsymbol{V}'$ are input of the {\MX}. The loss function of the {\MX} is:
\begin{equation}
    \mathcal{L}(\boldsymbol{V})=\frac{1}{L}\left(\boldsymbol{y}_{\boldsymbol{V}'}-\boldsymbol{V}\right)^2+1-\cos{\left(\boldsymbol{y}_{\boldsymbol{V}'}, \boldsymbol{V}\right)},\label{eq:MXLoss}
\end{equation}
where $\boldsymbol{y}_{\boldsymbol{V}'}$ is the reconstruction output of the {\MX} and $L$ is the length of it. Output of {MX}'s encoder is the final embeddings $\boldsymbol{E}$. It is noteworthy that only channel-mask policy is applied when generating embeddings for subsequent tasks. The embeddings have the same length as input, i.e.\ $nd=\dim \boldsymbol{V}=\dim \boldsymbol{E}$.

Fig. \ref{StrucMX} depicts the structure of the {\MX}. The encoder and the decoder are identical. They first expand the dimension to $4nd$, and then reduce it back to $nd$. Unlike the {\CMP}s, the {\MX} is not an undercomplete autoencoder. However, the noising process make sure that the inputs and the reconstruction targets of the {\MX} are different. In other words, identity mapping is naturally not an expected result of training the {\MX}. Therefore, the {\MX} does not fall into the trivial solution and will always produce meaningful outcomes if converges.

\subsection{Applying {\SYS} in particular tasks}
Pre-training a {\SYS} involves no particular labelled data. To implement {\SYS}, the \textit{pre-training} procedures are as follows:
\begin{enumerate}
    \item Select the processing method set $F=\left\{f_i|i\in[1,n]\right\}$.
    \item Collect an unlabeled dataset $U$ containing signal samples from the same type of sensors as the particular tasks. Apply the methods in $F$ for transformed datasets $\{f_i(U)\}$.
    \item Train $n$ {\CMP}s with transformed datasets separately. Apply {\CMP}s to transformed datasets for coded datasets $\{\boldsymbol{v}^i_{U}\}$. Concatenate the coded datasets to gather the concatenation dataset $\boldsymbol{V}_U$.
    \item Train a {\MX} with the concatenation dataset.
\end{enumerate}
To apply {\SYS} in a particular task $S$ about an annotated dataset $D_S^o$, researchers only have to:
\begin{enumerate}[resume]
    \item \label{step5} Apply the {\MX} to the annotated dataset $D_S^o$ for embeddings $\SYS(\hat{F}, D_S^o)$ with user-specific channel masks. The embeddings are the final outcomes of {\SYS}. Gather evaluation performance for all $\hat{F}\in \mathcal{P}(F)$ to find the best-performing method combination $F'$.
\end{enumerate}
For users, {\SYS} doesn't participate in additional training because there is no need for fine-tuning. Instead, a pre-trained {\SYS} simply offers embeddings of annotated data with user-specific channel masks. If researchers want to evaluate various configurations, the only work is to generate embeddings with corresponding masks. During this process, {\SYS} is fixed without the need for adjustment. Moreover, a trained {\SYS} can be utilized repeatedly since only step (\ref{step5}) involves particular datasets. Researchers can use it to generate wanted embeddings across tasks without any extra pre-training.
In contrast, knowledge-based traditional solutions require experts to manually design and test countless statistical features provided by processing methods, while data-driven non-reusable deep learning based models need repetitious training to adapt to different transformed datasets.

\section{Evaluation}\label{sec:evaluation}
We evaluate {\SYS} with acoustic and Wi-Fi signals. It is noteworthy that we don't necessarily compare with SOTA approaches in all experiments. The reason is that the experiments are designed to demonstrate {\SYS}'s general capability under different situations and perspectives, other than solely for acheiveing the best performance.
To be more specific, Acoustic experiments show {\SYS} can achieve comparable performance as traditional approaches, while Wi-Fi experiments are to show {\SYS}'s potential to work with other complicated models and their embeddings to achieve outstanding performance.

\subsection{Experiments on Acoustic Sensing}
\subsubsection{Datasets} We use \textbf{ESC-US} and \textbf{ESC-50} \cite{esc} for the evaluation. ESC-50 dataset comprises of 2,000 short environmental recordings split equally among 50 classes. 
ESC-US is an additional dataset including 250,000 recordings without labels. We use ESC-US as the unlabeled dataset for pre-training and ESC-50 for the downstream classification task.
\subsubsection{Evaluation Metrics and Baselines}
\cite{soundnet} evaluates the test accuracy of applying MFCCs with a convolutional network on the ESC-50 dataset so we use it as the metric. Although more recent works on this task have reported much better performance, they generally apply heavy neural networks and other information modalities, severly deviating from our evaluation aim, so we will not use their results.
\subsubsection{Implementation Details}
{\SYS} is implemented using Python and PyTorch \cite{pytorch}. It is trained in a PC with 1 NVIDIA GeForce RTX 3060 GPU, 32 GB memory, and an Intel(R) Core(TM) i5-11500 2.70GHz CPU. 
For each clip, we generate the log Mel-spectrogram which has two dimensions: frequency and time. We split the result along the frequency axis into four parts, each of which occupies a quarter of the whole band. We then reduce each part with addition, resulting in four sequences. We use the unlabeled dataset to train a prototype. The length of the codes is set to 128. Each {\CMP} is trained for 50 epochs and the {\MX} is trained for 100 epochs. The learning rates are 0.001 for both {\CMP}s and {\MX}, while the batch sizes are 64 and 128 respectively.
We use the same strategy as the baseline to evaluate the performance. We use {\SYS}-XXXX to denote the usage of four channels. The channel order is from the lowest frequency sequence to the highest. For example, {\SYS}-TFFF indicates that only the channel of the lowest frequency band is used.

\subsubsection{Performance}

\begin{figure}[htbp]
    \centering
    \includegraphics[width=0.5\columnwidth]{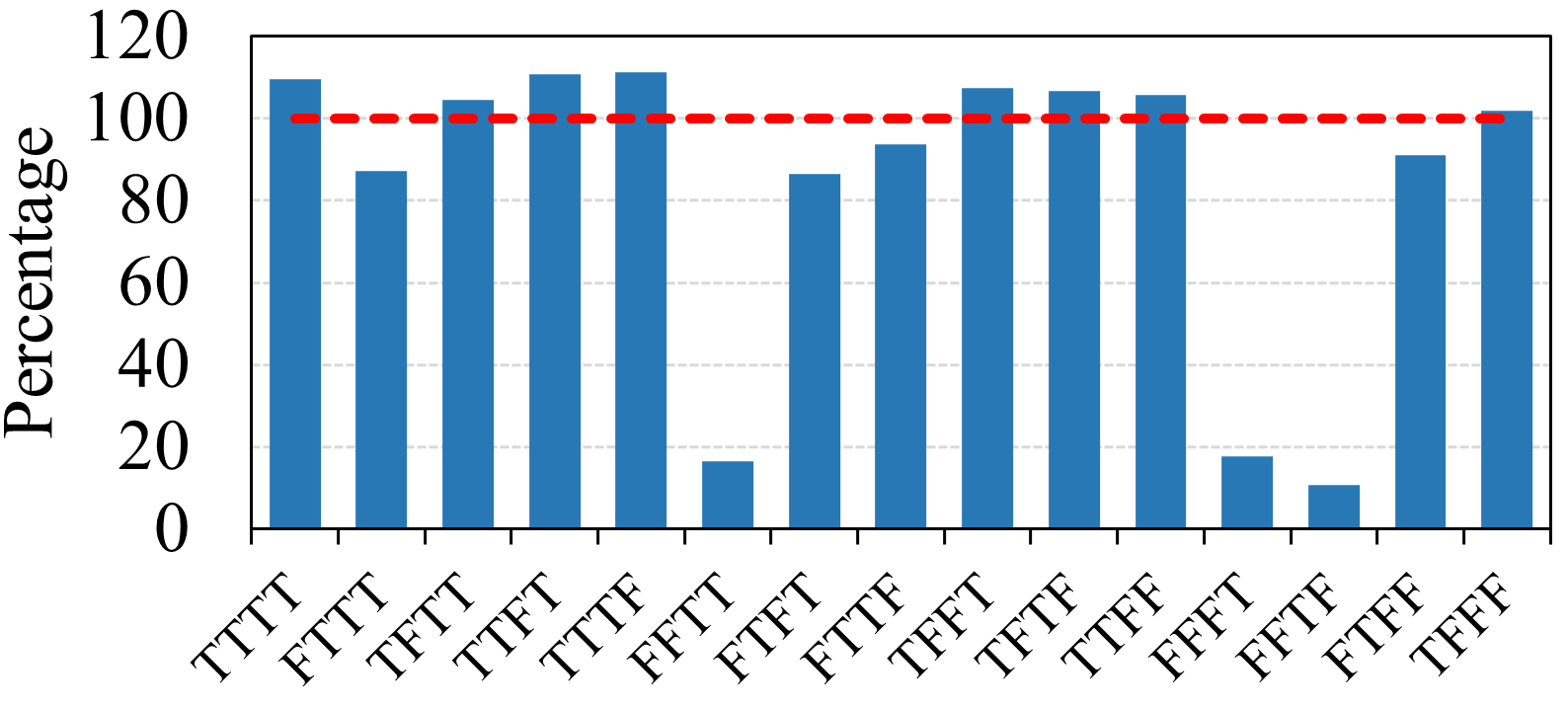}
    \caption{Ratio of {\SYS}'s accuracy to baseline's. The red line represents the baseline (100\%).}
    \label{fig:EvalAco}
\end{figure}

Ratio of {\SYS}'s accuracy to baseline's is depicted in Fig. \ref{fig:EvalAco}, where the red line represents the baseline (100\%). This evaluation is aimed to show that {\SYS} can achieve a comparable result with a commonly acknowledged method. It is depicted in Fig. \ref{fig:EvalAco} that the optimal performance among all mask configurations reaches similar classification accuracy with the baseline. Besides, we also find that the fourth channel, i.e. the highest frequency sequence, reports a relatively poor performance. Sound clips in ESC-50 include many human and animal sounds, where the sound frequencies are lower and their features are clearer. The environmental noises that have a higher frequency, however, are generally more chaotic. This might explain the performance difference between the low and high frequency channels.

\begin{figure*}[htbp]
    \centering
    \begin{subfigure}
        \centering
        \includegraphics[width=0.4\textwidth]{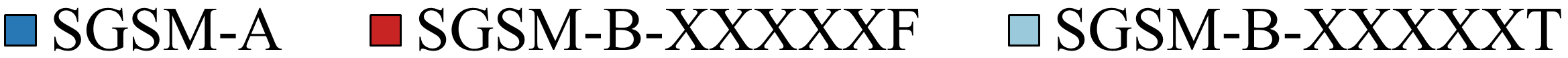}
    \end{subfigure}\\
    \begin{subfigure}
        \centering
        \includegraphics[width=\textwidth]{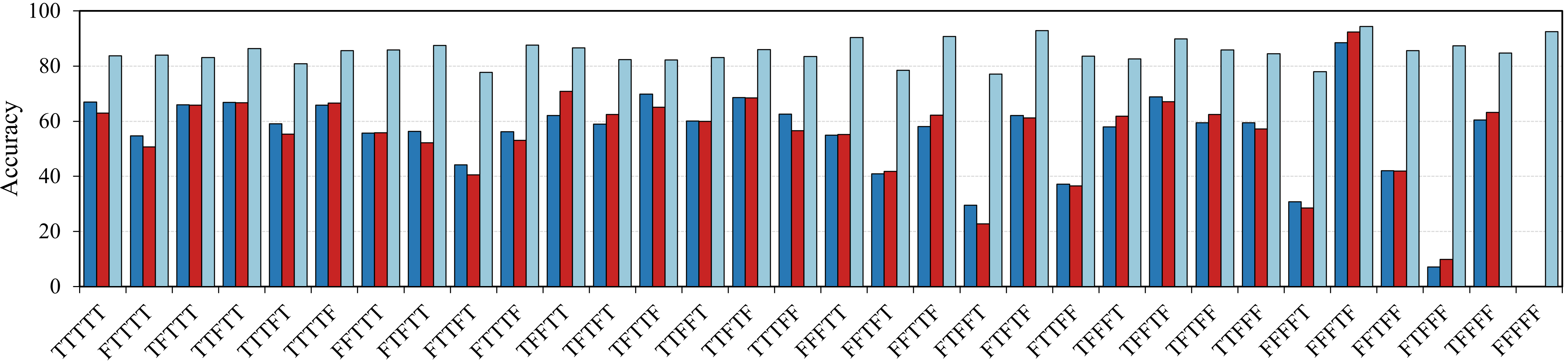}
    \end{subfigure}
    \caption{Performance of different method combinations. Combining simple methods and AutoFi results in better performance.}
    \label{fig:EvalWi-Fi}
\end{figure*}

\subsection{Experiments on Wi-Fi Sensing}

\subsubsection{Datasets} \textbf{NTU-Fi} \cite{ntuhar, ntuhuman} is a dataset that includes both human activity recognition (HAR) and human identification (Human ID) tasks. The data has a high resolution of subcarriers (114 per pair of antennas). Both datasets are separated into training and testing datasets. In this evaluation, we use the whole NTU-Fi HAR as the unlabeled dataset, the NTU-Fi Human-ID training dataset as the subsequent task training dataset, and the NTU-Fi Human-ID testing dataset as the subsequent task testing dataset.

\subsubsection{Evaluation Metrics and Baselines}
\textbf{AutoFi} \cite{autofi} is a Wi-Fi sensing model based on a self-supervised learning algorithm. It can generate an embedding for each Wi-Fi CSI sample. 
We train an instance of Auto-Fi with MLP backbone on NTU-Fi HAR without complicated parameter adjustments and fine-tuning. Although the model doesn't achieve the performance reported in \cite{autofi}, it is enough for us to conduct our evaluation. Since the evaluation metric is the testing accuracy in \cite{autofi}, we decide to use it as well.

\subsubsection{Implementation Details} In this evaluation, we test {\SYS}'s ability to combine simple processing methods and complex embeddings. We build two prototypes of {\SYS}. The hardware setup is the same as the one in acoustic signal experiments.
The first prototype, denoted as {\SYS}-A, has five channels whose signal processing methods are: DFT, DWT, Raw, HHT, and Periodogram. We treat each CSI sample as a combination of 342 signal sequences, corresponding to the 342 channels.
We let each {\CMP} generates a code of length 128, resulting in the final embeddings being in the shape of $(342, 128\times5)$. The second prototype, denoted as {\SYS}-B, has six channels, where the extra channel comes from the embeddings of AutoFi. We train an extra {\CMP} with AutoFi embeddings and a new {\MX} to combine all six channels. Thus, {\SYS}-B generates embeddings in the shape of $(342, 128\times6)$.

For evaluation, we first test AutoFi on NTU-Fi Human-ID. Since the embeddings are 1-dimensional, we use an MLP network including three fully connected layers as the classifier. Then, {\SYS}-A and {\SYS}-B use a convolutional network as the classifier because of the extra dimension. 
The classifiers are adjusted to having comparable sizes. All classifiers are trained on the NTU-Fi Human-ID training dataset for 150 epochs, and the accuracy are tested on the testing dataset. To denote results with different masks, we use {\SYS}-A-XXXXX and {\SYS}-B-XXXXXX. The highest five bits of both correspond to DFT, DWT, Raw, HHT, and Periodogram, respectively. The lowest bit for {\SYS}-B corresponds to the AutoFi channel.

\begin{table}[htbp]
\centering
\caption{Accuracy of AutoFi, {\SYS}-A, and {\SYS}-B}
\label{tab:WifiAcc}
\begin{tabular}{lr}
\toprule
Method       & Accuracy \\ \midrule
AutoFi       & 74.33    \\ 
{\SYS}-B-FFFFFT & 92.44    \\ 
{\SYS}-A-TTTTT & 67.00    \\ 
{\SYS}-B-TTTTTT & 83.71    \\ 
{\SYS}-A-FFFTF & 88.48    \\ 
{\SYS}-B-FFFTFT & 94.37    \\ \bottomrule
\end{tabular}
\end{table}
\subsubsection{Performance}
Classification performance is depicted in Fig. \ref{fig:EvalWi-Fi} and Tab. \ref{tab:WifiAcc}. In Fig. \ref{fig:EvalWi-Fi}, we group the accuracy results by the mask configurations of the first five channels. For example, group TTFTT includes accuracy of {\SYS}-A-TTFTT, {\SYS}-B-TTFTTF, and {\SYS}-B-TTFTTT. In group FFFFF, only performance of {\SYS}-B-FFFFFT is provided, because masking all channels is meaningless. In theory, combining information from five simple methods with the other AutoFi embeddings should result in a similar, if not better, performance compared to using them separately. In Tab. \ref{tab:WifiAcc}, we can see that {\SYS}-B-TTTTTT, i.e., combining codes from all channels, outperforms AutoFi and {\SYS}-A-TTTTT. This indicates that {\SYS}-B successfully incorporates information from codes of simple methods and codes of complicated embeddings. 

It is noteworthy that most masking combinations in {\SYS}-A report a better performance in {\SYS}-B with the sixth channel of AutoFi, as depicted in Fig. \ref{fig:EvalWi-Fi}. This indicates that AutoFi do capture features of the signals that the five simple methods don't. Besides, we notice that the masking combination with the best performance is not opening all channels, but the one opening HHT and AutoFi. It is possible that this combination extracts the most useful features of NTU-Fi Human-ID. Finally, it is worth mentioning that {\SYS}-B-FFFFFT performs significantly better than AutoFi alone. This implies that the {\MX} manages to discover latent information of AutoFi embeddings with the aid of other method channels and recover them for subsequent classification. Such results further demonstrate {\MX}'s effectiveness.

\section{Conclusion}\label{sec:conclusion}
This paper introduces {\SYS}, a new paradigm for sensing model. When compared to typical systems, {\SYS} can tackle a variety of tasks semi-automatically using less task-specific labelled data. 
Experiments conducted on two heterogeneous sensors show that {\SYS} works in various conditions, proving its broad applicability. Acoustic experiments show {\SYS} can achieve comparable performance as traditional approaches. Wi-Fi evaluations demonstrate that applying {\SYS} to an existing sensing model improves accuracy by 20\%.

\bibliographystyle{unsrt}  
\bibliography{references}

\end{document}